\documentclass[showpacs,eqsecnum,prd]{revtex4}
\usepackage[dvips]{graphicx}
\usepackage{amsmath}
\usepackage{amsfonts}
\usepackage{booktabs}
\newcommand{\Sw}{Schwarzschild }
\DeclareMathOperator{\erf}{erf}
\begin{document}

\title{Spacetime Foam Model of the \Sw Horizon}

\author{Jarmo M\"akel\"a} 
\email[Electronic address: ]{jarmo.makela@phys.jyu.fi}  
\author{Ari Peltola}
\email[Electronic address: ]{ari.peltola@phys.jyu.fi} 
\affiliation{Department of Physics, University of Jyv\"askyl\"a, PB 35 (YFL), FIN-40351 Jyv\"askyl\"a, Finland}

\begin{abstract}
We consider a spacetime foam model of the \Sw horizon, where the horizon consists of Planck size black holes. According to our model the entropy of the \Sw
black hole is proportional to the area of its event horizon. It is possible to express geometrical arguments to the effect that the constant 
of proportionality is, in natural units, equal to one quarter.
\end{abstract}


\maketitle

\section{Introduction}
In the inspired final chapter of their classic book Misner, Thorne and Wheeler state that there are three levels of gravitational collapse. The first two of them
are the gravitational collapse of the whole universe during the final stages of its recontraction, and the gravitational collapse of a star when a black hole
is formed. The third level of gravitational collapse is the quantum fluctuation of spacetime geometry at the Planck scale of distances. To rephrase Misner,
Thorne and Wheeler, ``collapse at the Planck scale of distances is taking place everywhere and all the time in quantum fluctuations in the geometry and,
one believes, the topology of space. In this sense, collapse is continually being done and undone...'' \cite{mtw}.

The picture given by Misner, Thorne and Wheeler of the Planck scale physics in these sentences is very charming. It immediately brings into one's mind
mental images of tiny wormholes and black holes furiously bubbling as a sort of spacetime foam. This is indeed a wonderful picture but unfortunatelly the 
ideas it suggests have never been taken very far. Instead of developing models of spacetime where spacetime consists of tiny wormholes and black holes the
reseachers in the field of quantum gravity have gone into another direction and developed, among other things, a very succesfull theory of loop quantum
gravity which treats spacetime as a spin network constituted of tiny loops.

The purpose of this paper is, in a certain sense, to revive some aspects of the old picture of spacetime as a foam of tiny wormholes and black holes.
Actually, there are reasons to believe that at the Planck length scale microscopic black holes might indeed play a role in the structure of
spacetime. For instance, suppose that we want to localize a particle within a cube with an edge length equal to one Planck length
\begin{equation} l_{\text{Pl}} := \sqrt{\frac{\hbar G}{c^3}} \approx 1.6 \cdot 10^{-35}\, \text{m}.
\end{equation}
It follows from Heisenberg's uncertainty principle that in this case the momentum of the particle has an uncertainty $\Delta p \sim \hbar / l_{\text{Pl}}$. 
In the ultrarelativistic limit the uncertainty in the energy of the particle is $\Delta E \sim c\Delta p$, which is around one Planck energy
\begin{equation} E_{\text{Pl}} := \sqrt{\frac{\hbar c^5}{G}} \approx 2.0 \cdot 10^{9}\, \text{J}.
\end{equation}
In other words, we have enclosed one Planck energy inside a cube whose diameter is one Planck length. That amount of energy, however, is enough to shrink the
spacetime region bounded by the cube into a black hole with a \Sw radius equal to around one Planck length. So it seems possible that when
probing the structure of spacetime at the Planck length scale one encounters with Planck size black holes.

At the present stage of research a construction of a precise mathematical model of spacetime as a whole made of tiny black holes is out of reach. However,
it is possible to test the idea of spacetime as a foam of microscopic black holes in the context of an important specific problem of quantum gravity. The
problem in question is the microscopic origin of black hole entropy.
There is a general agreement between the researchers working in the field of quantum gravity that black hole has an entropy which is one quarter of its event
horizon area or, in SI units,
\begin{equation} \label{eq:kili} S=\frac{1}{4}\frac{k_B c^3}{\hbar G}A.
\end{equation}
This result is sometimes known as the Bekenstein-Hawking entropy law.
In addition to black holes, it is valid for cosmological, de Sitter and Rindler horizons as well, and it seems to imply that in addition to the three
classical, there is an enormous amount of quantum-mechanical degrees of freedom in black holes. It is reasonable to expect that these additional degrees of 
freedom lie at the horizon of the black hole. During the recent ten years or so several attempts have been made by experts of string theory and loop quantum
gravity to identify the quantum-mechanical degrees of freedom of black holes, and to provide a microscopic explanation for the black hole entropy \cite{ash}.

To test the viability of an idea of spacetime as a foam of Planck size black holes we postulate, in this paper, a specific spacetime foam model of the event
horizon of a \Sw black hole \cite{gar}. It follows from our model that the entropy of the \Sw black hole is proportional
to its horizon area, and we also present arguments suggesting that the constant of proportionality must be, in natural units, equal to one quarter. In other
words, it seems possible to obtain Eq. (\ref{eq:kili}) from our model.

This paper is organized as follows: In section 2 we introduce our model, especially the postulates on which it is based. In short, the event horizon of a
macroscopic \Sw black hole is assumed to consist of Planck size, independent black holes. Each of the microscopic holes on the horizon is assumed to obey
a sort of ``Schr\"odinger equation'' of black holes, which was published some years ago. The postulates of our model connect the quantum states of individual 
microscopic holes on the horizon with the area eigenvalues of the event horizon of the macroscopic \Sw black hole.

In section 3 we proceed to calculate the entropy of the \Sw black hole from the postulates of our model. An essential ingredient of our calculation of black
hole entropy is our decision to apply \textit{classical} statistics to our model. This decision of ours is motivated by Hawking's result that, at least in the
semiclassical limit, black hole radiation spectrum is the continuous black body spectrum. If one attempted to apply any sort of quantum statistics to our model,
a discrete radiation spectrum radically different from the one predicted by Hawking would follow even for macroscopic black holes. Continuous spectrum,
however, is regained if classical, instead of quantum, statistics is applied to our model. We find that our model supplemented with classical statistics
implies that in a very low temperature the entropy of the \Sw black hole is, up to an additive constant, proportional to its horizon area. In section 4 it
is claimed, on grounds of certain geometrical arguments, that it is natural to take the constant of proportionality to be equal to one quarter. This result 
reproduces the Bekenstein-Hawking entropy law of Eq. (\ref{eq:kili}).

Section 5 contains a critical analysis of our model. We list the reasonable objections against our model we have managed to find, and our answers to these
objections.

\section{The Model}
The starting point of our model is Bekenstein's proposal of the year 1974, which states that the eigenvalues of the area of the event horizon of a black hole are
of the form:
\begin{equation} \label{eq:A}A_n = n\gamma l_{\text{Pl}},
\end{equation}
where $n$ is an integer and $\gamma$ is a pure number of order one. In other words, Bekenstein's proposal states that the event horizon area of a black hole 
has an equal spacing in its spectrum \cite{bek}.

Bekenstein's proposal has been revived by several authors on various grounds \cite{bmm}. One way to obtain Eq. (\ref{eq:A}) for \Sw black 
holes is to consider the 
following eigenvalue equation for the \Sw mass $M$ of the \Sw black hole (unless otherwise stated, we shall always use the natural units where 
$\hbar = c=G=k_B=1$):
\begin{equation} \label{eq:massa} \Bigg[ -\frac{1}{2}a^{-s} \frac{d}{da} \Bigg( a^{s-1}\frac{d}{da} \Bigg) +\frac{1}{2}a \Bigg] \psi(a) =M\psi(a).
\end{equation}
This equation which, in a certain very restricted sense, may be viewed as a sort of ``Schr\"odinger equation'' of the black hole from the point of view of a
distant observer at rest with respect to the hole, was obtained in Ref. \cite{lm} from the first principles by means of  an analysis which was based
on Kucha\v r's investigation \cite{kuc} of the Hamiltonian dynamics of \Sw spacetimes. 
In Eq. (\ref{eq:massa}), $a$ is the throat radius of the Einstein-Rosen wormhole
in a foliation where the time coordinate at the throat is the proper time of an observer in a free fall through the bifurcation two-sphere. $\psi(a)$ is
the wave function of the hole, and the real number $s$ determines the inner product between the black hole states such that 
\begin{equation} \label{eq:kemi}\langle \psi_1 |\psi_2 \rangle := \int\limits_0^\infty \psi_1^*(a) \psi_2(a)a^s da.
\end{equation}
In other words, the Hilbert space is chosen to be the space $L^2(\mathbb{R}^+,a^sda)$

It was shown in Ref. \cite{lm} that the spectrum of $M$ given by Eq. (\ref{eq:massa}) is always discrete, bounded from below, and can be made positive by means
of an appropriate choice of the self-adjoint extension. Moreover, it was shown that if $s=2$, then for large $n$ the WKB eigenvalues of the mass $M$ are
of the form
\begin{equation} \label{eq:vili} M_n^{\text{WKB}} = \sqrt{2n+1}.
\end{equation}
Surprisingly, this WKB approximation for large $n$ provides an excellent approximation for the mass eigenvalues even when $n$ is small (\textit{i.e.} of
order one): At the ground state where $n=0$, the result given by Eq. (\ref{eq:vili}) differs around one percent from the real ground state
mass eigenvalue, and the difference very rapidly goes to zero when $n$ increases (see Appendix A). So we see that because the event horizon area of a \Sw black
hole with mass $M$ is:
\begin{equation} \label{eq:isoA} A=16\pi M^2,
\end{equation}
Eq. (\ref{eq:massa}) implies, as an excellent approximation, the following spectrum for the horizon area:
\begin{equation} \label{eq:fili} A_n = 32\pi \Big( n+\frac{1}{2} \Big).
\end{equation}
In other words, we recover Bekenstein's proposal of Eq. (\ref{eq:A}) with $\gamma=32\pi$.

We are now prepared to state the postulates of our model. Our postulates are:
\begin{enumerate}
\item The event horizon of a macroscopic \Sw black hole consists of independent, Planck size \Sw black holes.
\item Each hole on the horizon obeys Eq. (\ref{eq:massa}) with $s=2$.
\item Holes in the ground state where $n=0$ do not contribute to the horizon area.
\item When the hole is in the $n$th excited state, it contributes to the horizon an area which is proportional to $n$.
\item The total area of the horizon is proportional to the sum of the areas contributed by the holes on the horizon.
\end{enumerate}

No doubt it is rather daring to assume that the microscopic holes on the horizon are independent of each other. However, at this stage of research that
assumption is necessary if we want to make progress. Moreover, one may speculate with a possibility that when the density of black holes on the horizon is
constant, the effects of all the rest of the holes to an individual hole on the horizon may cancel each other.
Another bold assumption is contained in the Postulate 3. As such, the Postulate 3 views observable black
holes as excitations of black holes in a ground state. When compared to the Postulates 1 and 3, the Postulates 2, 4 and 5 appear as rather natural. The
Postulates 4 and 5 imply that the horizon area is of the form
\begin{equation} \label{eq:jees} A=\alpha (n_1+n_2+ \ldots +n_N),
\end{equation}
where $n_1,n_2, \ldots,n_N$ are the quantum numbers associated with the area eigenstates of the holes on the horizon, and $\alpha$ is an unknown constant
of proportionality. We shall consider the postulates of our model in more details in section 5.

\section{Entropy}
When the concept of entropy of a black hole was introduced by Bekenstein and Hawking 30 years ago, all the derivations of the expression (\ref{eq:kili})
for black hole entropy were performed by means of \textit{semiclassical} arguments: When Hawking obtained an expression $T=\frac{1}{8\pi M}$ for black hole
temperature yielding the expression (\ref{eq:kili}) for black hole entropy, he treated spacetime classically and matter fields quantum-mechanically \cite{haw}. 
Later, in 1977, 
when Gibbons and Hawking calculated the black hole entropy by means of Euclidean path integral methods \cite{haw2}, 
they simply used a semiclassical approximation to the path integral.
In other words, the expression (\ref{eq:kili}) for the black hole entropy is, to the greatest possible extent, a ``semiclassical entropy'' which we should
obtain from a microscopic model of spacetime.

At this point it is useful to recall what we actually mean by the very concept of entropy in quantum and classical physics: In quantum statistics
the entropy of a system in a given macrostate is the natural logarithm of the number of microstates corresponding to that macrostate, whereas in classical
statistics entropy is the natural logarithm of the phase space volume corresponding to the given macrostate \cite{lan}. 
It may be shown that the entropy of quantum 
statistics reduces to the entropy of classical statistics when the spectra of observables are not assumed to be discrete but continuous. Which 
statistics --- classical or quantum --- should we use for spacetime itself?

The answer to this question depends on whether we consider the radiation spectrum of a black hole continuous or discrete. 
If the radiation spectrum is discrete, then the horizon area spectrum of the macroscopic hole, as well as the spectra of the microscopic holes on the horizon, 
is discrete and we must use quantum statistics. Consequently, if the radiation spectrum is continuous, the area spectrum is also continuous and we must 
use classical statistics. According to Hawking, the radiation spectrum is purely thermal, continuous 
black-body spectrum. Therefore the answer to our question is obvious: If we want to obtain for the black hole entropy an expression which is in agreement 
with Hawking's radiation law we must, of course, use \textit{classical} statistics for the black hole itself.

This conclusion has nothing to do with the question about whether we really believe black hole radiation spectrum to be continuous, as in Hawking's theory, or
discrete as it is, for instance, according to Bekenstein's proposal. Curiously, it is possible to express an argument, based on Heisenberg's uncertainty
principle, to the effect that when the effects of matter fields are so strong that they overshadow the quantum effects of spacetime, then the discrete spectrum
predicted by Bekenstein's proposal reduces to Hawking's black body spectrum \cite{mak}. 
Actually, this is what we assume here. More precisely, we assume that when the
effects of matter fields are absent, the black hole area spectrum is discrete and follows Bekenstein's proposal but when the effects of matter fields are
strong enough, the spectrum becomes continuous, and classical statistics may be applied to the black hole itself.

So, let us calculate the classical entropy, or the natural logarithm of the classical phase space volume, of our model. 
Eq. (\ref{eq:massa}) may be deduced from the classical Hamiltonian
\begin{equation} H=\frac{1}{2a}\big( p^2+a^2\big) ,
\end{equation}
of the \Sw black hole (For the details see Ref. \cite{lm}). In this equation $p$ is the canonical momentum conjugate to $a$. 
The numerical value of $H$ is the \Sw mass $M$ of the hole. Therefore we find, using Eq. (\ref{eq:vili}), that for a single black hole with ``mass'' $M_1$
and canonical coordinates $a_1$ and $p_1$ on the horizon,
\begin{equation} \label{eq:ralli} \frac{1}{4a_1^2}\big( p_1^2+a_1^2 \big)^2 = M_1^2 =2n_1 +1.
\end{equation} 
However, since we are now considering classical statistics, $n_1$ no more is an integer but it may be any non-negative real number. Moreover, it is no more
possible to associate with $M_1$ any sensible physical meaning as the ``mass'' of a hole on the horizon (A more detailed investigation of this issue is 
performed in section 5). $M_1$ simply is a parameter with the property
that $M_1^2$ is proportional to the area contributed by a single microscopic hole to the total horizon area.

Eq. (\ref{eq:ralli}) now implies that
\begin{equation} \label{eq:yka}\frac{1}{8a_1^2}\big( p_1^2+a_1^2 \big)^2 + \ldots +\frac{1}{8a_N^2}\big( p_N^2+a_N^2 \big)^2 = \frac{N}{2}+\frac{A}{\alpha}.
\end{equation}
It can be shown (See Appendix B) that for fixed $A$ and $N$ this is a compact $(2N-1)$-dimensional hypersurface in a $2N$-dimensional phase space. Its
$(2N-1)$-volume, in the units of $(2\pi \hbar)^{N-1/2}$, is of the form 
\begin{equation} \label{eq:risto} \Omega = C(N) \Bigg( N+\frac{2A}{\alpha}\Bigg)^{N-1/2},
\end{equation}
where $C(N)$ is a function which depends on $N$ only. For fixed $N$, $C(N)$ is finite.

At this point we should pause for a moment. In section 2 we threw away with great pomp and ceremony the vacuum contribution to the black hole horizon area by 
stating that holes in the ground state do not contribute to the horizon area. Eq. (\ref{eq:yka}), however, involves the term $\frac{N}{2}$ which 
is equal to the vacuum contribution to the horizon area. Are we now quietly returning back the vacuum contribution to our calculations?

The answer to this question is emphatetically no. The reason why the term $\frac{N}{2}$ suddenly appears in Eq. (\ref{eq:yka}) is that in Eq. 
(\ref{eq:yka}) we calculate the \textit{phase space volume} of our model, and the role of the term $\frac{N}{2}$ is to take into account the vacuum contribution
to the phase space volume. In other words, although we assume that ground states do not contribute to the horizon area, they nevertheless contribute to the
phase space volume. Actually, as we shall see in a moment, in very low temperatures most of the phase space volume is really occupied by the unobservable vacuum,
or black holes in the ground state. 

The entropy of the \Sw black hole is the natural logarithm of the phase space volume $\Omega$ corresponding to a fixed horizon area $A$:
\begin{equation} \label{eq:kiisseli} S = \ln \Omega = \ln C(N) + \Bigg( N-\frac{1}{2}\Bigg) \ln \Bigg( N +\frac{2A}{\alpha} \Bigg).
\end{equation} 
Using Eq. (\ref{eq:isoA}) and keeping $N$ as a constant we may obtain the temperature $T$ of the \Sw black hole:
\begin{equation} \frac{1}{T} = \Bigg( \frac{\partial S}{\partial E} \Bigg)_{N,V} = \frac{N-\frac{1}{2}}{N} \frac{64\pi M}{\alpha + \frac{32\pi M^2}{N}}
\end{equation}
or, since $N$ is assumed to be very large,
\begin{equation} \label{eq:T} T = \frac{\alpha}{64\pi M} +\frac{1}{2N}M.
\end{equation}
It is interesting that if $N$, the number of microscopic holes on the horizon, is sufficiently small when compared to the mass $M$ of the macroscopic hole
then the temperature increases, instead of decreasing as in Hawking's theory of black hole radiation, as a function of $M$.

At this point we introduce a new assumption to our theory. We assume that \textit{most microscopic holes on the horizon are in the ground state}, where $n=0$. 
More precisely, we assume that the average
\begin{equation} \bar{n}:= \frac{n_1+\ldots +n_N}{N}=\frac{A}{N\alpha}
\end{equation}
of the quantum numbers $n_1, \cdots, n_N$ has the property 
\begin{equation} \label{eq:oletus} \bar{n} \ll 1.
\end{equation}
Since $A$ is proportional to $M^2$, we find that this condition may be written as 
\begin{equation} \frac{M}{N} \ll \frac{1}{M}.
\end{equation}
When the assumption (\ref{eq:oletus}) is employed, the second term on the right hand side of Eq. (\ref{eq:T}) vanishes, and the temperature becomes to:
\begin{equation} \label{eq:T2} T = \frac{\alpha}{64\pi M}.
\end{equation}
The assumption (\ref{eq:oletus}) is sensible when we consider the thermodynamics of a macroscopic black hole in a \textit{very low temperature}: In a
very low temperature the constituents of any system tend to be as close to the ground state as possible. The same is true also for our model: We see that the
minimum temperature is achieved when the second term on the right hand side of Eq. (\ref{eq:T}), and therefore $\bar{n}$, becomes as small as possible.

From Eq. (\ref{eq:T2}) it now follows that, up to an additive constant which depends only on the number $N$ of the microscopic holes on the horizon, the 
entropy of the \Sw black hole is, in a very low temperature,
\begin{equation} \label{eq:elli} S=\frac{2A}{\alpha}.
\end{equation}
In other words we have obtained, up to an undetermined constant of proportionality, the Bekenstein-Hawking entropy law of Eq. (\ref{eq:kili}). Here we have 
considered \Sw horizon only but our consideration may easily be generalized for other horizons as well.

\section{Constant of Proportionality}
It only remains to fix the constant $\alpha$ in Eq. (\ref{eq:elli}). The constant $\alpha$ was defined in Eq. (\ref{eq:jees}). From that definition it
follows that if we know how the total area of the horizon depends on the areas of the holes on the horizon, we may calculate $\alpha$. A comparison of
Eqs. (\ref{eq:kili}) and (\ref{eq:elli}) yields the result that, in natural units, we should have $\alpha =8$, and our task is to obtain this value by means
of geometrical arguments. Of course, nobody really knows what the spacetime geometry at the Planck length scale really looks like, and therefore one should
take such arguments with a pinch of salt. However, it is possible to express arguments which, when not understood as a real description of spacetime structure
at the Planck length scale but rather as a sort of statistical average of the behaviour of the microscopic holes on the horizon, 
may at least provide a heuristic aid of thought.

\begin{figure}[htb!]
\begin{center}
\includegraphics{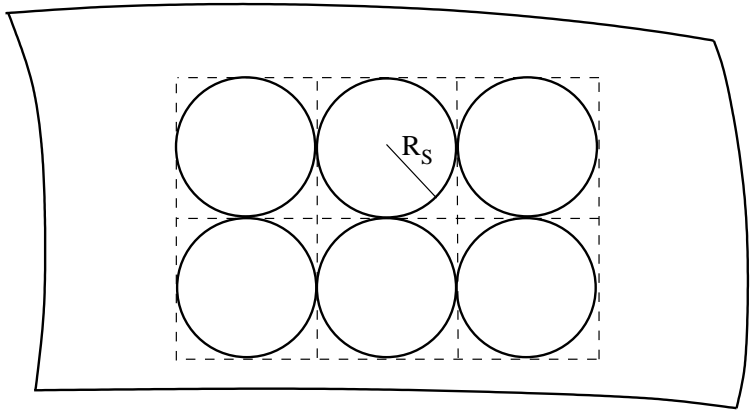}
\caption{In our model spacetime is constructed from Planck size black holes. A faraway observer sees each particular hole as a circle
with a \Sw radius $R_S$. When a hole on the horizon exactly fits inside a square, the area of the square is $\frac{4}{\pi}$ times the area of the
corresponding circle.\label{fig:neliot}}
\end{center}
\end{figure}

To begin with, consider Fig. \ref{fig:neliot}. In that figure we have drawn microscopic black holes on the horizon such that, at this stage, they do not
overlap each other. A distant observer does not see a hole on the horizon as a sphere but as a \textit{circle} with radius equal to its \Sw radius $R_S$.
The area of this circle is $\pi R_S^2$, which is one quarter of the horizon area of the corresponding microscopic hole. Eq. (\ref{eq:fili}), together with
the postulate 3, therefore implies that the area covered by a microscopic hole in the $n$th excited state is $8\pi n$.

When considering Fig. \ref{fig:neliot}, however, one finds that the holes on the horizon do not cover the whole horizon but certain voids remain. A better
way to cover the horizon is to use squares drawn around the holes such that each hole exactly fits inside the corresponding square. In that case the area of 
each square is $\frac{4}{\pi}$ times the area of the correponding circle. Summing the areas of the squares we find that the horizon area is 
\begin{equation} A=32(n_1 + n_2 +\ldots + n_N),
\end{equation}
from which it follows that $\alpha =32$. In other words, $\alpha$ is too large with a factor of 4.

A remedy for this problem may be found if we assume that the holes on the horizon overlap each other. Recall that we are obtaining an expression for the
maximum entropy of a macroscopic \Sw black hole. The maximum entropy may be gained if there is a maximum amount of microscopic holes on the horizon.
That is because if two black holes on the horizon come together to form a single black hole, the area of the resulting black hole is larger than is the sum
of the areas of the original black holes. In other words, a larger amount of the horizon is covered by a smaller number of degrees of freedom. Since the maximum
entropy is achieved when the number of degrees of freedom is as large as possible, it follows that when two black holes on the horizon overlap in the state
of maximum entropy, they still must not form a single black hole. 

\begin{figure}[htb!]
\begin{center}
\includegraphics{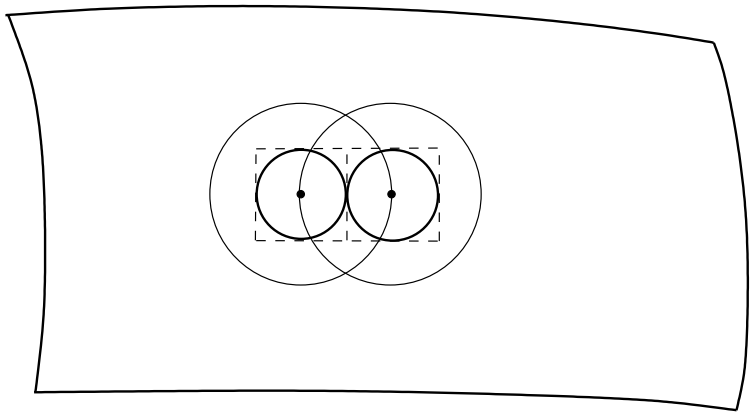}
\caption{Two identical black holes on the horizon overlapping each other such that the center points of the holes lie on each other's event horizons.
Covering the horizon with black holes overlapping each other in this manner is equivalent to covering the horizon with non-overlapping holes
having radiuses which are one halves of the radiuses of the original holes.\label{fig:overlap}}
\end{center}
\end{figure}

The maximum overlapping between two holes with equal sizes is achieved when the center points of the holes lie on each other's event horizons (See Fig. 
\ref{fig:overlap}). In that case the holes are not yet ``swallowed'' by each other, and the distance between the center points of the holes is their common 
\Sw radius
$R_S$. As one can see from Fig. \ref{fig:overlap}, however, covering the horizon with black holes overlapping each other in this manner is equivalent
to covering the horizon with holes with radius $\frac{1}{2} R_S$ and not overlapping each other. Therefore one finds that the horizon area now takes
the form:
\begin{equation} A = 8(n_1+n_2+ \ldots +n_N),
\end{equation}
which implies the desired result
\begin{equation} \alpha=8.
\end{equation}
When this result is substituted in Eq. (\ref{eq:elli}), we get
\begin{equation} S = \frac{1}{4}A,
\end{equation}
or, in SI units,
\begin{equation} S=\frac{1}{4}\frac{k_B c^3}{\hbar G}A.
\end{equation}
In other words, we have exactly recovered the Bekenstein-Hawking entropy of Eq. (\ref{eq:kili}).

However, it must again be strongly emphasized that the geometrical arguments given here should not be considered as a real description of the structure 
of spacetime
at the Planck length scale. Rather, they should be viewed as mere heuristic aids of thought. The microscopic structure of spacetime is certainly very much
more complicated than is the simple picture given in this section. Our geometrical arguments, however, have a certain statistical content which may be 
expressed as follows: The average number density of the microscopic black holes on the horizon is such that, if the holes were understood as classical objects,
their center points would lie, in average, on each other's event horizons. Whether one accepts this argument or not is, at the present stage of research,
a matter of taste but nevertheless it produces the desired value for the  constant $\alpha$.

\section{Some Objections}
As it may have become clear to the reader, our model and the postulates on which it is based, contain several ideas which
are physically new. Always when expressing a new physical idea it is of vital importance to try to prove oneself wrong, 
and this is what we shall attempt to do in this section. More precisely, we list all the reasonable objections against our
model we have managed to find, and our answers to these objections. Our objections are:

\begin{enumerate}
\item It is very strange to think a black hole horizon of being made of Planck size black holes. If black holes are made of 
Planck size black holes, then what are the Planck size black holes made of? 

\textit{Answer}: In our model we have assumed the Planck size black holes described by Eq. (\ref{eq:massa}) to be 
fundamental constituents, or building blocks, of spacetime. So they are assumed to have no intrinsic structure. The idea 
that spacetime might be made of Planck size black holes in one way or another is supported, among other things, by the 
arguments based on Heisenberg's uncertainty principle expressed in the Introduction. So far we have no precise model of the
whole spacetime being made of Planck size black holes, and a construction of such a model should be viewed as a challenge 
for the future. We only have a model of a \Sw horizon and, as one can see, our model reproduces the Bekenstein-Hawking
entropy law.

\item Although the paper intends to provide a model of a \Sw horizon, an assumption that a horizon of spacetime is considered, is nowhere explicitly used in
the derivation of the Bekenstein-Hawking entropy law. Actually, it seems that arguments similar to those employed in this paper may be used to imply that
every spacelike two-surface of spacetime has an entropy which is proportional to its area, no matter whether that two-surface is a horizon or not. Is this
not in a contradiction with the known properties of gravitational entropy?

\textit{Answer}: It is true that an assumption that a horizon of spacetime is under scrutiny is nowhere used in this paper. The primary reason for our choice
to construct a model of a \Sw horizon is that with a \Sw horizon one may associate a well-defined concept of energy which, in turn, may be used when calculating
the temperature of the horizon. It is also true that one could argue by means of reasoning similar to that used in this paper that entropy is associated not only
with spacetime horizons but with every spacelike two-surface of spacetime. More precisely, it really seems that every finite spacelike two-surface, no matter
whether that two-surface is a horizon or not, possesses an entropy which, in natural units, is one quarter of its area.

Although astonishing, this conclusion, however, does not contradict with the known physics. Actually, there seems to be very good grounds to believe that every
piecewise smooth, spacelike two-surface of spacetime indeed carries a certain amount of entropy although that entropy produces observable physical effects
only for the observers having that two-surface as a horizon. Provided that one accepts the view held at least implicitly by many authors that horizon entropy
is due to the microscpic degrees of freedom of spacetime at the horizon this result is a direct consequence of the well-established result that any finite
part of a Rindler horizon has an entropy which, in natural units, is one quarter of the area of that part. The line of reasoning producing this conclusion may
be expressed as follows:
\begin{itemize}
\item[(a)] Each finite part of a Rindler horizon possesses an entropy which is one quarter of its area.
\item[(b)] Each finite spacelike two-plane is, from the point of view of an appropriate observer, a part of a Rindler horizon.
\item[(c)] Therefore, each finite spacelike two-plane possesses an entropy which is one quarter of its area.
\item[(d)] Each piecewise smooth spacelike two surface is a union of infinitesimal spacelike two-planes, each having an entropy equal to one quarter of its area.
\item[(e)] Therefore, each piecewise smooth spacelike two-surface possesses an entropy which is one quarter of its area.
\end{itemize}

In other words, an explicit assumption that one looks at the horizon is not needed when one attempts to obtain the Bekenstein-Hawking entropy law: There are
good grounds to believe that the Bekenstein-Hawking  entropy law holds for any piecewise smooth, spacelike two-surface. We hope to be able to consider this
subject in more details in the forthcoming papers.

\item How can the Planck size black holes on the horizon be independent of each other? Certainly there should be an 
extremely strong interaction between them.

\textit{Answer}: This is indeed what one might expect on classical grounds. However, we are now considering spacetime at
very small length scales where quantum gravity reigns, and all bets are on. It is possible that the interactions between
black holes effectively cancel each other in a bit similar way as the gravitational effects cancel each other from the 
point of view of an observer at the center of the Earth. This is an open question.

In addition to these rather general remarks it is possible to express a much more serious argument supporting the claim
that the microscopic black holes might indeed be independent of each other. This argument is based on Jacobson's
exremely important discovery that Einstein's equations may be viewed as \textit{thermodynamical equations of state} \cite{jac}. More
precisely, Jacobson showed that if one assumes that a local Rindler horizon always possesses an entropy which is one 
quarter of its area, then Einstein's equations follow from the first law of thermodynamics. So one is inclined to think
that Einstein's equations, with their prediction that macroscopic black holes attract each other, are probably not very
fundamental at all but they are merely a consequence from the statistics of the fundamental constituents of spacetime. There
is not necessarily any interaction between these fundamental constituents at the microscopic level, but their statistical
properties imply at the macroscopic level properties which might lead one to conclude an existence of a sort of effective
interaction. As a familiar example one may think of a classical ideal gas: Classical ideal gas has a certain pressure
which might lead one to conclude (incorrectly) that between the particles of the gas there is a repulsive interaction 
which prevents one to compress the gas. However, no such repulsive interaction really exists, and the pressure is simply
a consequence from the statistical mechanics of the gas. Perhaps something similar happens with microscopic black holes:
At the microscopic level there is no interaction between the holes but their statistical properties imply in a certain 
limit, among other things, Einstein's equations with all their predictions. Actually this may well be the case, because
it seems to us that our analysis could be generalized to show that not only black holes but also a local Rindler horizon
possesses an entropy which is one quarter of its area, and, as Jacobson has showed, a generalization of the Bekenstein-Hawking entropy
law for local Rindler horizons implies Einstein's equations. So it is possible that the gravitational
interaction between macroscopic black holes may be a direct consequence from the independence of microscopic black holes. 

\item Is it not absurd that black holes in the ground state where $n=0$ do not contribute to the horizon area? 

\textit{Answer}: This is probably the most serious objection against our model. However, the idea that vacuum states do not
produce measureable effects (except in very special circumstances) is not very unfamiliar in physics: In quantum field 
theories in flat spacetime the field is decomposed into its Fourier components, and the possible energies of each component are of the 
form:
\begin{equation} E_{\vec k ,n}= \Big(n +\frac{1}{2}\Big)\hbar \omega_{\vec k},
\end{equation}
where $n=0,1,2,\cdots$ and $\omega_{\vec k}$ is the angular frequency of the component corresponding to the wave vector 
$\vec k$. The vacuum energy where $n=0$, is never observed but only the energies which are above the vacuum energy may be 
measured. In other words, everything one can observe are excitations of the vacuum but the vacuum itself cannot be
observed (except by means of the Casimir effect). It is not entirely impossible that this feature of nature might pertain
also to quantum gravity. In quantum gravity, however, energy is replaced by area: As one can see from Eq. (\ref{eq:fili}),
the possible areas of the microscopic holes on the horizon are exactly of the same form as are the energies of the  Fourier
components of a quantized field. One is therefore tempted to draw an analogy between the quantization of energy
in ordinary quantum field theories, and the quantization of area in our model: In the same way as everything one observes 
in ordinary quantum field theories are excitations of the vacuum state of energy, in our model everything one observes 
are excitations of the vacuum state of area. This analogy may be viewed as the main 
justification for the Postulate 3.

\item It follows from the Postulate 5 that the squares $M_i^2$ of the masses of the individual holes on the horizon sum
up, essentially, to the square $M^2$ of the total mass of the hole. Is this not wrong because certainly the masses 
themselves should sum up to the total mass of the hole?

\textit{Answer}: The concept of mass is very problematic already in classical general relativity. Actually, the concept of
mass may be properly defined only when spacetime is asymptotically flat, and in that case the relevant mass concept is
the ADM mass, which measures the mass-energy included by the whole spacetime. When we try to define the mass-energy 
included by a specific part of curved spacetime, however, we run into grave difficulties because the energy-momentum
tensor of the gravitational field, in general, is not well-defined. So it is not clear what we are talking about when
we say that the mass of a certain object in curved spacetime is such-and-such, unless that is the only object in an 
asymptotically flat spacetime, and we may define its ADM mass.

In particular, the concept of mass of an object at the event horizon of a black hole is very problematic. To see how 
problematic it really is, consider the four-momentum of a particle having a rest mass $m$ in flat spacetime:
\begin{equation} p^\mu =m\dot{x}^\mu,
\end{equation}
where the dot means proper time derivative. When spacetime is static, \textit{i.e.} the metric tensor does not depend on
the time parameter $x^0$, the quantity
\begin{equation} p_0 = m g_{0\mu}\dot{x}^\mu
\end{equation}
is conserved along geodesics and also along curves having timelike Killing vectors as their tangents, and it offers the best attainable definition 
for the concept of energy of a particle in curved
spacetime. Since the \Sw metric may be written as
\begin{equation} ds^2 = \Big( 1-\frac{2M}{r}\Big) dt^2 -\frac{dr^2}{1-\frac{2M}{r}} -r^2d\theta^2 -r^2 \sin^2 \theta 
\, d\phi^2,
\end{equation}
we find that
\begin{equation} p_0 =m\Big( 1-\frac{2M}{r}\Big)\dot{t}.
\end{equation}
For a particle at rest with respect to the \Sw coordinates we have
\begin{equation} \dot{t} =\frac{dt}{d\tau} =\Big( 1-\frac{2M}{r}\Big)^{-1/2},
\end{equation}
and therefore
\begin{equation} p_0 =m \Big( 1-\frac{2M}{r}\Big)^{1/2},
\end{equation}
Hence we see that when an object lies at the horizon where $r=2M$, its energy, according to above mentioned definition, is zero. 
Of course it is impossible to keep 
a particle still at the horizon but nevertheless our considerations provide an example of the problems of the mass-energy
concept at the event horizon of a black hole, even in the classical level. In the quantum level one may expect even deeper 
problems. Therefore, if we consider the event horizon of a black hole as a system made of tiny black holes, the mass
parameters $M_i$ of the holes do
not represent mass in any ordinary sense. Consequently, the sum of the parameters $M_i$ is not the \Sw mass $M$ of the macroscopic hole, 
either. However, each hole on the horizon covers a region with an area proportional to $M_i^2$ on the horizon. Therefore
it is natural to postulate that the sum of the quantities $M_i^2$ equals, up to a constant of proportionality, to the 
square $M^2$ of the \Sw mass of the hole.

\item The entropy of the hole has been calculated by using classical statistics right from the beginning. Would it not
have been more appropriate to perform the calculations by using quantum statistics first, and then take the classical
limit?

\textit{Answer}: This question has been considered in details already in section 3. Still, it is appropriate to repeat here the essential points:
It follows from Eq. (\ref{eq:jees}) which, in turn, is a direct consequence from our postulates, that the radiation spectrum of a black hole is discrete.
However, according to Hawking's radiation law the black hole radiation spectrum is a continuous black body spectrum. It is impossible to obtain Hawking's
continuous spectrum from our model unless one assumes right from the beginning that the quantum numbers $n_1, n_2, \cdots ,n_N$ may take any values,
and classical statistics applies. It follows
from Eq. (\ref{eq:jees}) that the possible frequences of the quanta of radiation are, for $\alpha =8$, integer multiples of the fundamental frequency
\begin{equation} v_0 := \frac{c^3}{8\pi^2 G} \frac{1}{M}.
\end{equation}
Even for macroscopic black holes (\textit{i.e.} when $M$ is of the order of ten solar masses) this quantity is fairly big: It is of the order of 0.3 kHz 
which is about the same as is the resolving power
of an ordinary portable radio receiver! So if one wants Hawking's continuous spectrum out from the discrete spectrum predicted by Eq. (\ref{eq:jees}),
the stationary quantum states of the black holes must become very much mixed with each other such that classical statistics, in effect, may be applied.
A mechanism for this kind of mixing through the interaction between the hole and matter fields was proposed in Ref. \cite{mak}.

\item The expression of Eq. (\ref{eq:kiisseli}) for the black hole entropy is not exactly one quarter of the horizon area, even for big $N$, but it contains
an additive constant which depends only on $N$. The presence of that term should produce physical effects different from those obtainable from the 
Bekenstein-Hawking entropy law.

\textit{Answer}: Actually, the quantity $N$ represents the \textit{particle number} in our model, the ``particles'' being now the microscopic black holes on
the horizon. In all thermodynamical systems the physically observable thermodynamical quantities are related to those partial derivates of entropy
where the particle number $N$ is kept constant during the differentiation. For instance, the temperature $T$ of a system is the inverse of the partial
derivative of entropy $S$ with respect to energy $E$ such that the particle number $N$ and the volume $V$ of that system are kept constant:
\begin{equation} \frac{1}{T} := \Bigg( \frac{\partial S}{\partial E} \Bigg)_{V,N}.
\end{equation}
Moreover, the pressure $p$ of a system is its temperature times the partial derivative of entropy with respect to the volume $V$ such that energy $E$ and
particle number $N$ are kept constant:
\begin{equation} p := T\Bigg( \frac{\partial S}{\partial V} \Bigg)_{E,N}.
\end{equation}
The only quantity which in any thermodynamical system is related to the partial derivative of entropy with respect to the particle number $N$ is the
\textit{chemical potential} 
\begin{equation} \mu := -T\Bigg( \frac{\partial S}{\partial N} \Bigg)_{E,V}
\end{equation}
of the system. As far as one is not interested in the chemical potential of the \Sw black hole, the physical predictions obtainable from our expression
for black hole entropy are identical to those obtainable from the Bekenstein-Hawking entropy law. As it comes to the chemical potential of the \Sw black
hole, in turn, nobody has the slightest idea about what that might be for the very simple reason that nobody knows what the ``particles'' constituting a
black hole would be. So we see that our expression for the black hole entropy does not contradict with the existing knowledge of black hole thermodynamics.
In general, it seems possible, at least in principle, to use different chemical potentials to distinguish different derivations of the 
Bekenstein-Hawking entropy law. More precisely, even if a certain derivation of the Bekenstein-Hawking entropy law yielded an expression for the temperature
of the horizon consistent with that law, the number of degenerate states corresponding to the same horizon area $A$ is not necessarily $e^{\frac{1}{4}A}$
but an expression for the number of degenerate states may contain a model-dependent prefactor, which is a function of the chemical potential of the system.

\end{enumerate}

\section{Concluding remarks}
In this paper we have considered a spacetime foam model of the \Sw horizon where the horizon of a macroscopic \Sw black hole consists of Planck size \Sw
black holes. Using this model we found that the entropy of a macroscopic \Sw black hole is, up to an additive constant,  proportional to its 
horizon area. It seems that
our derivation of this result is valid for any horizon but here we have restricted our consideration to \Sw horizons only. Certain heuristic arguments
may be employed to imply that the constant of proportionality is, in natural units, equal to one quarter.

The key points in our derivation were an assumption that when a microscopic black hole on the horizon is in a ground state, it does not contribute to the 
horizon area, and our decision to apply \textit{classical} statistics to our spacetime foam model. An assumption that a hole in a ground state, even if its
\Sw mass were non-zero, does not contribute to the horizon area, may be motivated by an analogy with ordinary quantum field theories: In ordinary
quantum field theories the vacuum state of energy is not observed, whereas in our model the vacuum state of energy is replaced by the vacuum state of area.
As it comes to our
decision to apply classical, instead of quantum statistics to our model, that decision may be considered justified on grounds of the fact that according to 
Hawking the radiation spectrum of a black hole is a continous black body spectrum. If we want to obtain a continuous spectrum for that radiation we must
assume that the spectra of observables of a macroscopic black hole are continuous, and therefore we must use classical statistics for the black hole itself.

In our model we assumed at first that the tiny holes on the horizon obey Bekenstein's proposal. In other words, we assumed that when the effects of matter 
fields are neglected, the horizon area spectra of the holes on the horizon have an equal spacing. That spacing was obtained from Eq. (\ref{eq:massa}),
the ``Schr\"odinger equation'' of the \Sw black hole. When the effects of matter fields are assumed to be strong enough, however, the discrete area spectrum
washes out into continuum, and Hawking's continuous black body spectrum is recovered. In other words, we considered the implications of Eq. 
(\ref{eq:massa}) in a limit where the uncertainties in the area eigenvalues are of the same order of magnitude as is the spacing between the area eigenvalues
of nearby states. The fact that we used Eq. (\ref{eq:massa}) in our derivation of an expression for the black hole entropy may be viewed as an
argument supporting the physical validity of that equation as well as of Bekenstein's proposal. In this paper we do not express opinions about whether
the radiation spectrum really is continuous or discrete. We only showed that if the radiation spectrum is assumed to be continuous, then the 
Bekenstein-Hawking entropy law follows from our model.

Taken as a whole, our model may be viewed as an attempt to understand the microscopic structure of spacetime. 
No doubt the picture provided by our model is very different
from those provided by, for instance, string theory and loop quantum gravity. An andvantage of our model, however, is that it takes seriously the possibility
suggested, among other things, by Heisenberg's uncertainty principle, that spacetime at the Planck length scale might consist of Planck size black holes.
So far we have no precise model of spacetime itself but only of its \Sw horizon. However, our results with the derivation of the Bekenstein-Hawking entropy law
are encouraging, and it will be very interesting to see whether the ideas presented in this paper may be worked out into a precise mathematical model of the 
microscopic structure of spacetime.

\acknowledgements
We thank Matias Aunola for invaluable assistance in the numeric analysis of the mass spectrum given by Eq. (\ref{eq:liitu}). We also thank Jorma Louko
and Markku Lehto, as well as an unknown referee, for their constructive criticism during the preparation of this paper.

\appendix

\section{Mass Eigenvalues}
During the preparation of this paper we have studied the mass eigenvalues of Eq. (\ref{eq:massa}), when $s=2$, for small $n$ in two different ways. 
The first method we have used is perturbation theory. The second method is to solve Eq. (\ref{eq:massa}) numerically. 
Both of these methods, of which the second one is more reliable, seem to imply that Eq. (\ref{eq:vili}) provides an excellent approximation for the mass 
eigenvalues even for small $n$. 

\subsection{Perturbation Theoretic Approach}
As a starting point we consider Eq. (\ref{eq:massa}) when $s=2$:
\begin{equation} \label{eq:kukka} \Bigg[ -\frac{1}{2a}\Bigg( \frac{d^2}{da^2} +\frac{1}{a}\frac{d}{da} \Bigg) +\frac{1}{2}a\Bigg]\psi (a) = M\psi (a)
\end{equation}
We write the Hamiltonian of Eq. (\ref{eq:kukka}) in the form
\begin{equation} \hat{H} = -\frac{1}{2a} \frac{d^2}{da^2} - \frac{1}{2a^2}\frac{d}{da}+\frac{1}{2}a=: \hat{H}_0 +\hat{H}',
\end{equation}
where
\begin{equation} \hat{H}_0 := -\frac{1}{2a} \frac{d^2}{da^2}+\frac{1}{2}a
\end{equation}
and 
\begin{equation} \hat{H}' := - \frac{1}{2a^2}\frac{d}{da}.
\end{equation}
If we neglect the term $\hat{H}'$ in Eq. (\ref{eq:kukka}) and denote 
\begin{equation} u:=a-M,
\end{equation}
we see that the resulting differential equation takes the form:
\begin{equation} \label{eq:kari} \Big( -\frac{1}{2}\frac{d^2}{du^2}+\frac{1}{2}u^2 \Big)\psi (u) = \frac{1}{2}M^2 \psi (u).
\end{equation}
This equation is similar to that of a one-dimensional harmonic oscillator, and therefore members of one set of solutions to Eq. (\ref{eq:kari}) are of the form
\begin{equation} \label{eq:pasi} \psi_n^{(0)} (a) = N_n H_n(a-M_n^{(0)}) e^{-\frac{1}{2}(a-M_n^{(0)})^2},
\end{equation}
where $H_n$ denotes Hermite polynomial of order $n$, $N_n$ is a normalization constant, and the eigenvalues $M_n^{(0)}$ are
\begin{equation} \label{eq:iop}M_n^{(0)} = \sqrt{2n+1}.
\end{equation}
One can see that the eigenvalues of Eq. (\ref{eq:iop}) are the same as the WKB eigenvalues in Eq. (\ref{eq:vili}).

Now we consider the term $\hat{H}'$ as a small perturbation of the unperturbed Hamiltonian $\hat{H}_0$. Here we must use somehow ``non-standard''
methods for calculating the first order perturbation because the Hamiltonian $\hat{H}_0$ is not a hermitian operator with respect to the inner product 
(\ref{eq:kemi}). Usually, when using perturbation theory, one writes the Hamiltonian in the form $\hat{H}=\hat{H}_0 +\lambda \hat{H}'$, where the perturbation
parameter $\lambda \in [0,1]$. During the calculation of first order perturbation one must operate with $\hat{H}_0$ to the bra-vectors $\langle \psi_n^{(0)}|$. 
The problem is that if $\hat{H}_0$ is not a hermitian operator, 
it cannot be transported from the right to the left hand side inside the Dirac brackets,
and the first order perturbation cannot be calculated in a standard way. However, if we assume that our perturbation expansion is viable 
when $\lambda =1$, we can set right from the beginning that $\lambda =1$. In this case the 
operator $\hat{H}_0 + \lambda \hat{H}$ is indeed a hermitian operator and one can operate with this operator to the bra-vectors in the usual manner. 
Using this trick,
one can calculate the first order perturbation. Let us begin with the perturbation expansions
\begin{equation} |\Phi_n \rangle = |\psi_n^{(0)} \rangle +\lambda |\psi_n^{(1)} \rangle + \lambda^2 |\psi_n^{(2)} \rangle +\ldots,
\end{equation}
\begin{equation} M_n  = M_n^{(0)} +\lambda M_n^{(1)} + \lambda^2 M_n^{(2)} +\ldots,
\end{equation}
where the vectors $|\Phi_n \rangle$ are eigenstates of the Hamiltonian $\hat{H}$ and the numbers $M_n$ are the corresponding eigenvalues. Symbols 
$|\psi_n^{(k)}\rangle $ and
$M_n^{(k)}$ denote corrections of order $k$ to the unperturbed eigenstate $|\psi_n^{(0)} \rangle $ and to the eigenvalue $M_n^{(0)}$.
When these expansions are inserted into the equation $\langle \psi_n^{(0)} |\hat{H} |\Phi_n \rangle = \langle \psi_n^{(0)} |M_n |\Phi_n \rangle $, we get:
\begin{eqnarray} & & \langle \psi_n^{(0)}|(\hat{H}_0+\lambda \hat{H}') |\psi_n^{(0)} \rangle +
\lambda \langle \psi_n^{(0)}|(\hat{H}_0+\lambda \hat{H}')|\psi_n^{(1)} \rangle + 
\lambda^2 \langle \psi_n^{(0)}|(\hat{H}_0+\lambda \hat{H}') |\psi_n^{(2)} \rangle +\ldots = \\ \nonumber
& & \ \ \ \ \ \ \ \ \ \ \ \ \ \ \ \  \bigg( M_n^{(0)} +\lambda M_n^{(1)} + \lambda^2 M_n^{(2)} +\ldots \bigg) \bigg( \langle \psi_n^{(0)}|\psi_n^{(0)} \rangle +
\lambda \langle \psi_n^{(0)}|\psi_n^{(1)} \rangle + \lambda^2 \langle \psi_n^{(0)}|\psi_n^{(2)} \rangle +\ldots \bigg) \\ \nonumber
&\Leftrightarrow & \langle (\hat{H}_0+\lambda \hat{H}') \psi_n^{(0)} |\psi_n^{(0)} \rangle +
\lambda \langle (\hat{H}_0+\lambda \hat{H}') \psi_n^{(0)}|\psi_n^{(1)} \rangle + 
\lambda^2 \langle (\hat{H}_0+\lambda \hat{H}') \psi_n^{(0)}|\psi_n^{(2)} \rangle +\ldots = \\ \nonumber
& & \ \ \ \ \ \ \ \ \ \ \ \ \ \ \ \  \bigg( M_n^{(0)} +\lambda M_n^{(1)} + \lambda^2 M_n^{(2)} +\ldots \bigg) \bigg( 1 +
\lambda \langle \psi_n^{(0)}|\psi_n^{(1)} \rangle + \lambda^2 \langle \psi_n^{(0)}|\psi_n^{(2)} \rangle +\ldots \bigg) \\ \nonumber
&\Leftrightarrow &\lambda \bigg( \langle \psi_n^{(0)}| \hat{H}' |\psi_n^{(0)} \rangle -M_n^{(1)}\bigg) + 
\lambda^2 \bigg( \langle \hat{H}' \psi_n^{(0)}|\psi_n^{(1)} \rangle -M_n^{(1)} \langle \psi_n^{(0)}|\psi_n^{(1)} \rangle -M_n^{(2)} \bigg)+\ldots = 0.
\end{eqnarray}
Hence the first order perturbation is of the standard form
\begin{equation} M_n^{(1)} = \langle \psi_n^{(0)} | \hat{H}'|\psi_n^{(0)} \rangle.
\end{equation}

Let us now calculate the first order perturbation explicitly by using the definition (\ref{eq:kemi}) of the inner product. We get:
\begin{eqnarray} \label{qwe} M_n^{(1)} &=& \langle \psi_n^{(0)} | \hat{H}'|\psi_n^{(0)} \rangle = \int_0^\infty {\psi_n^{(0)}}^*(a) 
\Big( -\frac{1}{2a^2} \frac{d}{da} 
\psi_n^{(0)}(a) \Big)a^2\, da \nonumber \\
&=& -\frac{1}{2} N_n^2 \int_0^\infty H_n \big( a-M_n^{(0)} \big)\, e^{-(a-M_n^{(0)})^2} \Big[ H_n' \big( a-M_n^{(0)} \big) +
\big( M_n^{(0)} -a \big) \cdot H_n \big( a-M_n^{(0)} \big) \Big]da,
\end{eqnarray}
where $H_n' \big( a-M_n^{(0)} \big) = \frac{d}{da}H_n \big( a-M_n^{(0)} \big)$, and the coefficients $N_n$ are determined by 
the requirement
\begin{equation} \label{eq:hirvi} \int_0^\infty {\psi_n^{(0)}}^*(a) \psi_n^{(0)}(a) a^2\, da = 1.
\end{equation}
It turns out that the integral in Eq. (\ref{qwe}) can be written in a very simple form:
\begin{equation} \label{eq:lehma} M_n^{(1)} = \frac{1}{4}N_n^2 H_n^2 \big( M_n^{(0)} \big)\, e^{- {M_n^{(0)}}^2}.
\end{equation}
Integrals in Eq. (\ref{eq:hirvi}) can be solved analytically by means of the error function $\erf (x)$, but unfortunately a simple general formula 
for the coefficient $N_n$ seems to be somehow out of reach. However, one easily finds that
\begin{equation} N_0^2 = \Bigg[ \frac{1}{2}\, e^{-1} +\frac{3}{4} \sqrt{\pi } \big( 1 + \erf (1) \big) \Bigg]^{-1},
\end{equation}
and therefore, according to Eq. (\ref{eq:lehma}), we have
\begin{equation}  M_0^{(1)} = \frac{1}{4} \Bigg[ \frac{1}{2} +\frac{3}{4}\, e \sqrt{\pi } \big( 1 + \erf (1) \big) \Bigg]^{-1} \approx 0.0349228.
\end{equation}

First order perturbations in the cases where $n= 1, 2, 3, \ldots $ can also be evaluated analytically but integration in Eq. (\ref{eq:hirvi}) becomes more 
complicated when $n$ increases.
Therefore, one might find some mathematical programs (like Mathematica) as very useful tools when one tries to calculate the coefficients $N_n$ analytically. 
In Table \ref{tbl:tulokset} we give the numerical values of first order perturbations in the cases where $n=0,1, \ldots, 10 $.

However, the perturbation method has two weak points. Firstly, we cannot be sure that our perturbation expansion converges. We can only hope that first order
perturbations provide a good approximation to the difference between the real and the WKB eigenvalues of the \Sw mass $M$. 
Secondly, it is hard to find out to which self-adjoint extension our perturbation approach is connected 
(For details see Ref. \cite{lm}). For these reasons our results for first order perturbations can be thought only as suggestive and we need a more reliable
method for the study of the eigenvalues of $M$. This method will be discussed in the next section.   
\begin{table}[h!]
\centering
\begin{tabular}{llll} \hline
$\ \ n$ \ \ \ & $\ \ \ M_n^{(1)}$ & $\ \ \ M_n \ \ $ & $\ \ \ M_n^{\text{WKB}} \ \ $ \\ \cline{1-4}
\ \ 0 & $\ \ \ 0.0349228$ & $\ \ \ 1.01$ & $\ \ \ 1$ \\
\ \ 1 & $\ \ \ 0.0093795 \ \ \ \ \ $ & $\ \ \ 1.74\ \ \ $ & $\ \ \ 1.7320\ldots \ \ $ \\  \ \ 2 & $\ \ \ 0.00513602$  & $\ \ \ 2.24$ & $\ \ \ 2.2360\ldots $ \\  
\ \ 3 & $\ \ \ 0.00346$ & $\ \ \ 2.65$ & $\ \ \ 2.6457\ldots $ \\  \ \ 4 & $\ \ \ 0.00257746$ & $\ \ \ 3.01$ & $\ \ \ 3$ \\  
\ \ 5 & $\ \ \ 0.00203793$ & $\ \ \ 3.32$ & $\ \ \ 3.3166\ldots $ \\  \ \ 6 & $\ \ \ 0.00167622$ & $\ \ \ 3.61$ & $\ \ \ 3.6055\ldots $ \\  
\ \ 7 & $\ \ \ 0.001418$ & $\ \ \ 3.88$ & $\ \ \ 3.8729\ldots $ \\  \ \ 8 & $\ \ \ 0.00122503$ & $\ \ \ 4.13$ & $\ \ \ 4.1231\ldots $ \\  
\ \ 9 & $\ \ \ 0.00107574$ & $\ \ \ 4.36$ & $\ \ \ 4.3588\ldots $ \\  \ \ 10 & $\ \ \ 0.000957045$  & $\ \ \ 4.59$ & $\ \ \ 4.5825\ldots $ \\ \hline
\end{tabular}
\caption{Numerical values of $M_n^{(1)}$, $M_n$ and $M_n^{\text{WKB}}$. The results are given with the best possible precision our numerical analysis
may offer. One may observe that WKB eigenvalues for the mass $M$ provide an
excellent approximation to the exact mass eigenvalues even when $n$ is small.  \label{tbl:tulokset}}
\end{table}

\subsection{Numerical Analysis}
The numerical analysis of the mass eigenvalues has been performed with Femlab, and the results can be found in Table \ref{tbl:tulokset}. In our numerical 
analysis we solve eigenvalues for the differential equation
\begin{equation} \label{eq:liitu} \Bigg( -\frac{9}{8} \frac{d^2}{dx^2} -\frac{9}{32} \frac{1}{x^2}+\frac{1}{2} x^{2/3}\Bigg) \psi (x) =M \psi (x)
\end{equation}
with the boundary conditions $\psi (0) = \psi (\infty )=0$ (Of course, the boundary condition $\psi (\infty )=0$ must be put in as $\psi (a)=0$,
where $a$ is a ``sufficiently'' large number). It may be shown that the mass spectrum given by this equation is identical to the mass spectrum
given by Eq. (\ref{eq:kukka}) \cite{lm}. 
One can also study which self-adjoint extension the numerical routine has chosen when calculating the mass eigenvalues: The eigenfunctions
corresponding to $M_n$ in Table \ref{tbl:tulokset} seem to behave, within the limits of the precision of the numerical computing, 
like $\psi (x) \sim \sqrt{x}$ when $x \to 0$. This observation fixes the self-adjoint extension for our eigenvalues.

Finally, it should also be noted that, due to a limited accuracy of numerical computing, some of the eigenvalues $M_n$ might be around one percent 
smaller than the values in Table \ref{tbl:tulokset}. This does not cause any problems and actually the mass eigenvalues are brought even 
closer to the WKB estimate of Eq. (\ref{eq:vili}).

\section{Phase Space Volume}
In our model, phase space volume corresponding to a fixed horizon area $A$ is determined by the condition (\ref{eq:yka}):
\begin{equation} \label{eq:yup}\frac{1}{8a_1^2}\big( p_1^2+a_1^2 \big)^2 + \ldots +\frac{1}{8a_N^2}\big( p_N^2+a_N^2 \big)^2 = \frac{N}{2}+\frac{A}{\alpha}.
\end{equation}
This equation describes a closed, compact $(2N-1)$-dimensional hypersurface $\Sigma$ in the $2N$-dimensional phase space spanned by 
the canonical coordinates $a_i$ and $p_i$.
To calculate its volume we choose coordinates $\lambda_1, \cdots, \lambda_{N-1} \in [0,1]$ and 
$\varphi_1, \cdots ,\varphi_N \in [0,2\pi ]$ on the hypersurface $\Sigma$ such that
\begin{eqnarray} & & \label{eq:koor1} a_1=L(\lambda_1+\lambda_1 \cos \varphi_1), \\ & &p_1=L\lambda_1 \sin \varphi_1, \\ 
& & a_2=L(\lambda_2+\lambda_2 \cos \varphi_2), \\ & &p_2=L\lambda_2 \sin \varphi_2, \\ 
& & \vdots \nonumber \\ 
& & a_N = L(\lambda_N + \lambda_N \cos \varphi_N), \\
& & \label{eq:koorN} p_N =L\lambda_N \sin \varphi_N.
\end{eqnarray}
Here $L = \sqrt{N+\frac{2A}{\alpha}}$, and $\lambda_N = \lambda_N(\lambda_1, \cdots , \lambda_{N-1})$ can be calculated from Eq. (\ref{eq:yup}):
If one substitutes the coordinates $a_i$ and $p_i$ in terms of $\lambda_1, \cdots, \lambda_{N}$ and 
$\varphi_1, \cdots ,\varphi_N$ into Eq. (\ref{eq:yup}), one gets
\begin{equation} \label{eq:uusi} \lambda_N = \sqrt{1-\lambda_1^2- \lambda_2^2- \ldots - \lambda_{N-1}^2}.
\end{equation}
Eq. (\ref{eq:uusi}) represents a hypersphere in the $N$-space spanned by the coordinates $\lambda_i$. So it is natural to perform another 
coordinate transformation
from the coordinates $\lambda_i$ to the generalized spherical coordinates $\theta_1 , \cdots , \theta_{N-2} \in [0,\frac{\pi}{2} ]$ and 
$\phi \in [0,\frac{\pi}{2} ]$:
\begin{eqnarray} & & \label{eq:ym1}\lambda_1= \cos \phi \prod_{i=1}^{N-2} \sin \theta_i  ,\\
& &  \lambda_2= \sin \phi \prod_{i=1}^{N-2} \sin \theta_i  ,\\
& & \lambda_3=  \cos \theta_1 \prod_{i=1}^{N-3} \sin \theta_i ,\\
& &  \lambda_4= \cos \theta_2 \prod_{i=1}^{N-4} \sin \theta_i ,\\ & & \vdots \nonumber \\ 
& & \label{eq:ym2} \lambda_N = \cos \theta_{N-2}.
\end{eqnarray}

Generally, the volume of the any (smooth) $N$-dimensional hypersurface $S$ can be evaluated from the integral
\begin{equation} \int\limits_S \sqrt{g}\, d^N x,
\end{equation}
where $g$ is the determinant of the metric. In our case the position vector of a given point on the hypersurface $\Sigma$ can formally be written as
\begin{eqnarray} \vec r &=& a_1 (\varphi_1, \cdots ,\varphi_N, \theta_1, \cdots , \theta_{N-2},\phi)\, \hat{e}_1 
+p_1 (\varphi_1, \cdots ,\varphi_N, \theta_1, \cdots , \theta_{N-2},\phi)\, \hat{e}_2+ \ldots \\ \nonumber 
& & + a_N (\varphi_1, \cdots ,\varphi_N, \theta_1, \cdots , \theta_{N-2},\phi)\, \hat{e}_{N-1}
+p_N (\varphi_1, \cdots ,\varphi_N, \theta_1, \cdots , \theta_{N-2},\phi)\, \hat{e}_N,
\end{eqnarray}
where $\hat{e}_i$'s $(i=1, \cdots ,2N)$ are orthogonal unit vectors in our $2N$-dimensional phase space. 
From Eqs. (\ref{eq:koor1})-(\ref{eq:koorN}) 
it follows that all non-vanishing components of the metric on the hypersurface, defined by the relation
\begin{equation} g_{ij} = \vec b_i \cdot \vec b_j,
\end{equation}
where $\vec b_i = \frac{\partial \vec r}{\partial x^i}$ ($i=1, \cdots ,2N-1$) is the tangent vector of the coordinate curve corresponding to the
coordinate $x^i$, are proportional to  $L^2$, and otherwise contain only products of the sines and the cosines of the angles $\phi$, $\theta_i$ and $\varphi_i$. 
Therefore the 
phase space volume $\Omega$ corresponding to fixed area $A$ has, in the units of $(2\pi \hbar)^{N-1/2}$, the form:
\begin{equation} \label{eq:renki} \Omega = \frac{L^{2N-1}}{(2\pi )^{N-1/2}} 
\int\limits_0^{\pi /2}d\phi \int\limits_0^{\pi /2}d\theta_1 \cdots \int\limits_0^{\pi /2}d\theta_{N-2} 
\int\limits_0^{2\pi}d\varphi_1 \cdots \int\limits_0^{2\pi}d\varphi_N \sqrt{\tilde{g}},
\end{equation}
where $\tilde{g}$ is the determinant of the metric whose components $\tilde{g}_{ij}$ are obtained from the components $g_{ij}$ such that we simply divide
each $g_{ij}$ by $L^2$.

The $(2N-1)$-dimensional integral in Eq. (\ref{eq:renki}) is, unfortunately, very hard to calculate. However, one can easily see that this integral converges:
From Eqs. (\ref{eq:koor1})-(\ref{eq:koorN}) and (\ref{eq:ym1})-(\ref{eq:ym2}) it follows that $\tilde{g}$ is everywhere finite (it contains only products of
the sines and the cosines of the angles $\phi$, $\theta_i$ and $\varphi_i$). Furthermore, the integration is performed over a bounded region because
$\varphi_i \in [0,2\pi ]$ and $\theta_1,\phi \in [0, \frac{\pi}{2}]$. Therefore, the $(2N-1)$-dimensional integral in Eq. (\ref{eq:renki}) converges. 
Its value, however, depends only on $N$ and we can denote
\begin{equation} \frac{1}{(2\pi )^{N-1/2}} \int\limits_0^{\pi /2}d\phi \int\limits_0^{\pi /2}d\theta_1 \cdots \int\limits_0^{\pi /2}d\theta_{N-2} 
\int\limits_0^{2\pi}d\varphi_1 \cdots \int\limits_0^{2\pi}d\varphi_N \sqrt{\tilde{g}} =:C(N).
\end{equation}
So we can write the phase space volume $\Omega$ as:
\begin{equation} \Omega = C(N) \Bigg( N+\frac{2A}{\alpha} \Bigg)^{N-1/2},
\end{equation}
which is Eq. (\ref{eq:risto}).


\begin{thebibliography}{99}

\bibitem{mtw} C.~W.~Misner, K.~S.~Thorne and J.~A.~Wheeler, \textit{Gravitation} (W.~H.~Freeman and Company, New York, 1973).
\bibitem{ash} See, for example, A.~Ashtekar, J.~Baez, A.~Corichi and K.~Krasnov, Phys. Rev. Lett. \textbf{80} (1998) 904-907,
A.~Ashtekar, J.~Baez and K.~Krasnov, Adv. Theor. Math. Phys. \textbf{4} (2000), 1, A.~Strominger, and C.~Vafa, Phys. Lett. \textbf{B379} (1996) 99,
G.~T.~Horowitz in \textit{Black holes and Relativistic Stars},  
ed. by R.~M.~Wald (University of Chicago Press, 1998). A comprehensive review of the present status
of loop quantum gravity and string theory has recently been written by L.~Smolin, hep-th/0303185. See also T.~Padmanabhan and A. Patel, hep-th/0305165,
gr-qc/0309053.
\bibitem{gar} As itself, our idea to construct \Sw horizon from microscopic black holes is not quite new. Similar ideas have been expressed by R.~Garattini,
Phys. Lett. \textbf{B459} (1999) 461, Nucl. Phys. Proc. Suppl. \textbf{88} (2000) 297, Int. J. Phys. \textbf{D4} (2002) 635, Entropy \textbf{2} (2000) 26,
gr-qc/0111068. See also F.~Scardigli, Class. Quant. Grav. \textbf{14} (1997) 1781.
\bibitem{bek} J.~D.~Bekenstein, Lett. Nuovo Cimento, \textbf{11} (1974) 467.
\bibitem{bmm} See, for example, J.~D.~Bekenstein and V.~F.~Mukhanov, Phys. Lett. \textbf{B360} (1995) 7, 
P.~C.~Mazur, Phys. Rev. Lett. \textbf{57} (1987) 929, Y.~Peleg. Phys. Lett. 
\textbf{B356} (1995) 462, A.~Barvinsky and G.~Kunstatter,
Phys. Lett. \textbf{B389} (1996), 231, H.~A.~Kastrup, Phys. Lett. \textbf{B389} (1996) 75, S.~Hod, Phys. Rev. Lett. \textbf{81} (1998) 4293, S.~Hod,
Gen. Rel, Grav. \textbf{31} (1999) 1639, H.~A.~Kastrup, Ann. Phys. (Leibzig) \textbf{9} (2000) 503, M. Bojowald and H.~A.~Kastrup, Class. Quant. Grav.
\textbf{17} (2000) 3009, A.~Barwinsky, S.~Das and G.~Kunstatter, Phys. Lett. \textbf{B517} (2001) 415, C.~Vaz, Phys. Rev. \textbf{D61} (2000) 064017.
A.~Barvinsky, B.~Das and G.~Kunstatter, Class. Quant. Grav. \textbf{18} (2001) 4845, J.~D.~Bekenstein, Int. J. Mod. Phys \textbf{A17S1} (2002) 21,
J.~D.~Bekenstein and G.~Gour, Phys. Rev. \textbf{D66} (2002) 024005.
\bibitem{lm} J.~Louko and J.~M\"akel\"a, Phys. Rev. \textbf{D54} (1996) 4982. The results for \Sw black holes were later generalized for Reissner-Nordstr\"om
black holes by J.~M\"akel\"a and P.~Repo, Phys. Rev. \textbf{D57} (1998) 4899, and for Kerr-Newman black holes by J.~M\"akel\"a, P.~Repo, 
M.~Luomajoki and J.~Piilonen,
Phys. Rev. \textbf{D64} (2001) 024018. See also J.~M\"akel\"a, Found. of Phys. \textbf{32} (2002) 1809.
\bibitem{kuc} K.~Kucha\v r, Phys. Rev. \textbf{D50} (1994) 3961.
\bibitem{haw} S.~W.~Hawking, Commun. Math. Phys. \textbf{43} 1975 199.
\bibitem{haw2} G.~W.~Gibbons and S.~W.~Hawking, Phys. Rev. \textbf{D15}, 2752 (1977).
\bibitem{lan} See, for example, L.~D.~Landau and E.~M.~Lifshitz, \textit{Statistical Physics} (Butterworth-Heinemann, 1999).
\bibitem{mak} J.~M\"akel\"a, Phys. Lett. \textbf{B390} (1997) 115.
\bibitem{jac} T.~Jacobson, Phys. Rev. Lett. \textbf{75} (1995) 1260.

\end{thebibliography}
\end{document}